# ELSA as an asset to the EU Quantum Act

## A position paper

**Adrian Schmidt**[1*] **Zeki C. Seskir**[1] **Pieter Vermaas**[2]

*[1]Institute for Technology Assessment and Systems Analysis, Karlsruhe Institute of Technology, Germany*

*[2]Ethics and Philosophy of Technology Section, Delft University of Technology, The Netherlands*

## Executive summary

In this position paper, we present European expertise on the ethical, legal, and social aspects (ELSA) of quantum technologies as an asset for the EU Quantum Act. As Europe prepares the Act at a crucial stage in the development of quantum technologies (QT), it should build on Europe's distinctive strengths. Those include, alongside a strong research base, successful industrial players, and skilled talents, Europe's human-centered approach and its responsible development of technologies.

We therefore argue that ELSA expertise is essential to the successful development of quantum technologies in Europe. By translating European values into technological practices, governance structures, and concrete policy measures, it can contribute to making Europe's approach both responsible and competitive. ELSA expertise should therefore become a central part of the EU Quantum Act.

We present our position by means of eight recommendations. They are meant to showcase the opportunities that ELSA brings to the development of quantum technologies. The recommendations move beyond ex-post regulation and an exclusive risk-based assessment. Instead, we promote the integration of European values, social inclusion, and global cooperation as a truly European approach. They can leverage European strengths, contribute to a unique selling point for quantum technologies made in Europe, and support a truly sovereign technology development.

*Corresponding author: adrian.schmidt2@kit.edu

These eight recommendations are invitations to European policy makers. They serve their purpose if they inspire to strengthen the Quantum Act and European policy on quantum technologies by incorporating ELSA expertise:

### RECOMMENDATIONS

1. Set key European targets as moonshots central in quantum technologies.
2. Embed European values central in quantum technologies.
3. Align the development of quantum technologies with motivations of younger generations.
4. Emphasize international cooperation and trade as a core European strength.
5. Develop quantum technologies applications globally and inclusively.
6. Make technology assessment a continuous part of quantum technologies' development.
7. Create shared sociotechnical repositories.
8. Develop a European narrative for the future of quantum technologies.

To maximize its inspirational potential and facility their translation into policy, we call for an immediate exchange on ethical, legal and social aspects between the larger ELSA for QT community and policymakers involved in developing the Quantum Act. Such a session could be convened under the European Parliament's Panel for the Future of Science and Technology (STOA). It should become a forum for discussing how those recommendations and further ELSA expertise can be incorporated into the Act, enabling Europe to govern quantum technologies in a way that builds on its distinctive strengths.

## 01 Introduction: The European ELSA asset

Europe's established expertise in ELSA of emerging technologies provides a strategic foundation for shaping quantum sovereignty responsibly and should be embedded in the European Quantum Act.

Expertise in ELSA, that is, in the *Ethical, Legal and Social Aspects* of technologies, is a European asset. It is an asset by which Europe has shaped its research, industry and governance of recent emerging technologies, and it should do now for quantum technologies. *Privacy* is a defining element in the European *General Data Protection Regulation* (GDPR). *Human rights* are central to the European AI Act. *Responsible Research and Innovation* is the European basis for all its scientific and technological efforts. And it are ELSA experts who gave content to these values. For quantum technologies, *technological sovereignty* may become the main value that shapes European research, industry and governance. ELSA expertise is Europe's asset for giving meaning to this value commitment in the upcoming Europe Quantum Act, thus letting Europe realize its sovereignty in a responsible way.

We in the European quantum ecosystem need to do work to cease this opportunity. Currently ELSA and the European investments in the quantum domain are pitched as incompatible. Then ELSA is presented as slowing down the European development of quantum technologies, and these investments are in turn framed as societally blind (Coenen et al., 2022).

We should abandon this typecasting and adopt the European strength of including ELSA in technology development. As ELSA scholars, we take up this promise in this position paper, while recognizing that our views do not necessarily represent those of all ELSA scholars. We advance this promise taking inspiration from the role that concept cars play in the automotive industry. Car manufacturers develop and present concept cars to show their vision and their potential to customers and investors. Similarly, we present in this paper recommendations that may convince you of our vision and the potential of ELSA. Concepts cars are not necessarily taken into production yet shape the future; we similarly submit our recommendations for inspiration to Europe when drafting its Quantum Act.

## 02 The European quantum opportunity

Quantum technologies offer Europe a strategic opportunity to translate scientific excellence into lasting economic, political, and societal value, with the EU Quantum Act shaping the priorities, infrastructure, and governance choices that will determine how this potential is realized.

We have consensus about the potential that quantum technologies represents. Across computing, communication and sensing, quantum technologies may address complex computational problems more effectively, support secure digital communication, and enable more precise measurement and detection. QT is therefore not a narrow technological niche but rather an important enabling layer for future economic and social development.

For Europe, this makes quantum technologies significant not only as a field of scientific excellence, but also as a domain in which technological capability, infrastructure development, and strategic direction intersect. While many applications are still emerging, the future of QT remains open and can be shaped. Therefore, governance matters: choices made now may influence which applications are prioritized, which infrastructures are built, and which societal needs QT development will serve, with long-lasting consequences for technological and societal possibilities.

This holds especially for the upcoming EU Quantum Act. In a crucial time in the development of quantum technologies, the Act will define how QT are governed, funded, and shaped. It will contribute to reaping the economic, political, and social benefits of Europe's excellent quantum research (European Commission, 2025) and is therefore a vehicle to address shortcomings while incorporating European strengths.

## 03 | The importance for early-on ELSA integration

Europe's tradition of integrating ethical, legal, and social considerations is a strategic strength that can make quantum development more legitimate, resilient, and globally responsive while advancing technological sovereignty on distinctly European terms.

One of the strengths of Europe is it reliance on ethical, legal and social aspects in technological development. This strength is a response to the specific way Europe is organized. The European Union is not a single, socially homogeneous polity, but a common market and governance space composed of 27 Member States with different historical experiences, cultural backgrounds, and socio-economic models. Its political system is therefore shaped by balancing, mediation, and negotiated compromise. This early ELSA integration may make Europe's decision-making processes slow when compared to other entities such as the USA or China, but should not be understood as a weakness. It is one of Europe's structural strengths and an existential reality.
Recent debates on dependencies in digital technologies and Europe's insufficient digital sovereignty have sharpened this point. Federal structures, lengthy coordination processes, and administrative complexity may reduce speed, but they also contribute to stability and legitimacy. Europe should therefore not seek to compensate for these reflexive features simply by adapting governance approaches developed elsewhere. The more relevant task is to design governance in a way that makes European strengths effective under current geopolitical conditions.

A sovereign European approach should be truly based on European strength, adjusted to current global developments, and in a responsive manner to further changes in the next decade. In this context, sovereignty means not only independent technological capability, but also the ability to helping shape the terms on which technologies are developed and governed, as well as the ability to self-reflexively discuss what the sovereignty should lead to. Europe's focus on human needs, concerns, and expectations across a highly heterogeneous population therefore becomes a strategic asset. In this sense, the European objective of linking productivity growth with social inclusion, frequently mentioned in the Draghi report (European Commission: European Political Strategy Centre, 2025), requires practical operationalization rather than rhetorical endorsement.

This is where ELSA approaches are particularly important. They make societal concerns visible and governable by translating diffuse expectations, risks, and conflicts into forms that can be anticipatorily addressed in research, innovation, and policy. They are therefore not an external constraint on quantum development, but a strategic European resource. This includes the early identification and mitigation of risks as well as the amplification of benefits. As other technology fields have shown, the absence of early attention to societal concerns can contribute to resistance or backlash (Coenen & Grunwald, 2017). Early ELSA integration is thus not about obstructing innovation, but about observing concerns in time and incorporating them in ways that support socially robust and strategically sustainable development, and about developing technologies in a socially desirable and competitive way (OECD, 2024; Vermaas & Mans, 2024).

At the same time, these societal dimensions are not confined to a narrow European context. Quantum technologies depend on globally distributed developers, supporting institutions and users, and they will evolve within international markets and research ecosystems. Other regions are therefore not only potential recipients of European technologies, but also co-developers whose social contexts, institutional conditions, and expectations may shape technological trajectories. Early ELSA integration thus points beyond the articulation of European priorities alone and also concerns the relation between these priorities and broader global societal needs (OECD, 2025; Vermaas & Mans, 2024).

## 04 | The ongoing European ELSA research in quantum

Europe already possesses a growing and internationally connected ELSA research ecosystem that can help shape quantum technologies proactively, but the EU Quantum Act must integrate this expertise early and systematically as a core element of quantum governance.

A broad set of national, European and international actors has already recognized the importance of integrating ELSA perspectives into quantum technology development and has begun to articulate this in practice. National initiatives and organizations such as QDNL in the Netherlands (QDNL, 2019) and the TU Munich Quantum Social Lab in Germany[1], point towards formats in which high-level quantum research is linked with societal reflection rather than treated as separate domains. At the international level, institutions such as UNESCO and the Open Quantum Institute at CERN have also created spaces in which the wider needs, concerns and expectations associated with quantum development can be discussed across regions, as demonstrated in the International Year of Quantum in 2025 (UNESCO, 2026). At the same time, networks of leading ELSA researchers, including initiatives such as responsible quantum technologies (ResQT), indicate that a broader community is already in place to examine how regional differences shape the societal implications of QT and how these implications might be addressed constructively (Schmidt, Artaud, et al., 2025). Europe therefore does not need to build this perspective from scratch; it can draw on an existing and increasingly connected body of expertise (Vermaas & Mans, 2024).

Some of these actors have also made visible, in a specifically European context, that the strategic development of QT requires more systematic attention to ELSA dimensions. Technology assessment work, including the TAQT study by the German Institute for Technology Assessment and Systems Analysis (ITAS) of the Karlsruhe Institute of Technology (KIT) (Schmidt, Seskir, et al., 2025), as well as the study on the military use of QT (Grünwald et al., 2024), has emphasized the importance of continuously and concurrently assessing societal implications as a basis for proactive management rather than mere obstruction. This is significant because it clarifies that ELSA integration is not a secondary add-on once technological pathways have already solidified. It is part of the strategic capacity to shape those pathways in ways that are socially robust and politically sustainable.

[1] https://www.gov.sot.tum.de/innotech/research/projects/quantum-social-lab/

The key point, therefore, is whether it is being taken up early and consistently enough in European quantum policy. If the EU Quantum Act is to connect technological ambition with social inclusion in a credible way, existing ELSA capacities need to be treated less as peripheral consultation and more as part of the strategic architecture of quantum governance. It is precisely at this juncture that current European debates reveal both meaningful progress and still unresolved imbalances.

## 05 | The limits of ELSA integration

Current approaches to EU quantum policies risk narrowing ELSA integration to security, risk assessment, and ex-post regulation, requiring a renewed framework that responds to geopolitical and economic pressures while preserving societal legitimacy, sustainability, and inclusion.

The question, however, is how these considerations can be integrated into the current debate in a way that is effective, constructive and valuable for all actors involved. From the perspective of the authors, the present discussion around the EU Quantum Act already contains important elements, but it also reveals a number of remaining shortcomings.

One concern is that an understandable and necessary focus on security may lead to other equally relevant dimensions receiving less attention at this stage. These include societal embeddedness, sustainability or the long-term conditions for public acceptance. This could not only narrow the debate, but may also create disadvantages in the medium term, especially if quantum technologies become more deeply integrated into societal infrastructures. A second concern relates to secure supply chains and strategic autonomy. While both are central to Europe’s future position in QT, they may, if interpreted too narrowly, come at the expense of cooperation with important future markets and emerging partner regions. In such a case, these regions may increasingly orient themselves towards other global QT actors. A third concern is the calls for greater speed and efficiency in European QT development. While understandable in light of geopolitical competition and economic pressure, such calls may, when linked to deregulation, reduce societal aspects mainly to risk assessment and ex-post regulation, rather than treating them as proactive conditions for the long-term functioning and legitimacy of QT (Lewis, Adam M et al., 2025; OECD, 2025).

Taken together, the current constellation suggests a tension. In the context of a necessary acceleration and stronger strategic focus of European QT governance, ELSA aspects may move into the background. This could create medium- and long-term problems, including societal resistance or backlash. At the same time, it may also mean that broader societal benefits of QT are not fully realized, even where the technological potential exists.

As ELSA scholars, this also raises a self-critical question. If these insights have long been developed, discussed and published by a broader ELSA community, why do current European QT priorities appear to be shifting in a different direction? It seems that ELSA arguments may no longer be as persuasive under present political and geopolitical conditions than they were only a few years ago. To put it somewhat pointedly, the search for broadly integrated ELSA principles appears in some discussions to have been reduced to a focus on risk assessment and ex-post regulation (Kop, 2025). At the same time, we do not assume that questions of societal responsibility, inclusion and sustainability have simply become irrelevant to decision-makers. For this reason, the task is not only to repeat familiar arguments more loudly. It is also to reflect critically on how ELSA approaches are communicated, justified, and designed under changed conditions.

In what follows, we therefore seek to take a step towards decision-makers and formulate an alternative offer. This offer builds on the extensive and successful work already undertaken by the ELSA community, but it also rests on a critical reflection of existing approaches. It attempts to take seriously the changed circumstances of geopolitics, security risks and economic pressure. In this sense, it seeks to adapt ideas from ELSA research to a transformed world situation without making them arbitrary or reducing them to mere strategic vocabulary.

## 06 | The concept-car ELSA offer

The concept-car approach presents ELSA recommendations as forward-looking demonstrations of how European quantum governance could become more responsible, human-centered, and strategically aligned with European values, without requiring their immediate or comprehensive implementation.

Expertise in the ELSA of technologies is a European asset. It is an asset we want to showcase taking inspiration in the role that *concept cars* play in the automotive industry:

car manufacturers develop and present concept cars for showing their vision and their potential to customers and investors. We adopt this approach to be able to highlight the assets of ELSA for the development of QT, for according to the design literature concept cars are in the automotive industry not evolving into "*production models, but that is also not the intention. [Concept cars] are in fact simply the physical embodiment of an approach, a direction, a brand statement.*" (Gardien, 2006). And that is also our intention: showing the power of ELSA, while for the moment taking distance from implementation.

The recommendations we list, do not need to leave the showroom immediately. Rather, they are meant to demonstrate what could become possible if ELSA integration were pursued more systematically and long-term within European governance of QT. They illustrate what might follow if the long-standing insights of a broad and diverse ELSA community were used more consistently to support responsible and sustainable QT development in Europe. It points to the possibility of taking more seriously the concerns of many scholars, especially regarding longer-term risks that may be mitigated through early societal integration, as well as the chance to steer the development of QT in Europe in a more human-centered, socially, and economically beneficial way. Above all, it suggests that ELSA aspects can be understood not only as external constraints, but also as expressions of European values and, in this sense, as a strategic asset for the EU.

The following recommendations are therefore invitations. They seek to attract the attention of the policy maker who may want to move one of these concept vehicles from the showroom into the EU Quantum Act and beyond. And it is also an offer to the customers in Europe: to those who might want precisely this kind of vehicle, because it speaks to a vision that concerns them, that they can recognize themselves in and that may motivate them to join the journey towards a European quantum future. The recommendations sometimes overlap, so no need to adopt them all; if one already inspires you, they have served their purpose.

# 07 | Recommendations

## 1 | SET KEY EUROPEAN TARGETS AS MOONSHOTS CENTRAL IN QUANTUM TECHNOLOGIES

Europe should put its key policy targets as moonshots central into the Quantum Act. Since the Quantum Act will become the central legislative document for the development of quantum technologies by Europe and its Member States, it needs to align with EU's overall policy goals. Creating (technological) sovereignty and security are already central goals in the preparational documents of the Quantum Act,[2] the European Green Deal, the energy transition, the extension and maintenance of Europe's infrastructures, and aims as nuclear fusion, should become key targets for development and procurement of quantum technologies. When those targets are included as well, the Quantum Act will become more aligned with the EU's overall strategic priorities, therefore increasing legitimacy and policy consistency.

- **Quantum as the secure, sovereign and breakthrough tech**

## 2 | EMBED EUROPEAN VALUES CENTRAL IN QUANTUM TECHNOLOGIES

Europe should use the Quantum Act to embed its foundational and societal values systematically into quantum technologies development. These values include "prosperity, equity, freedom, peace and democracy in a sustainable environment".[3] Embedding these values is a way of shaping technological pathways in line with Europe's human-centered approach and its early integration of ethical, legal, and social considerations. At the same time, this process should be based on dialogue and consent rather than on the unilateral imposition of European values. It should also include female voices and currently marginalized groups, since the people involved are not only potential workforce or customers, but also central to an honest human-centered approach. Because quantum technologies are highly specialized and will evolve across interconnected markets, embedding of European values in this way will have broader positive impact.

- **Quantum as the societally relevant tech**

---

[2] The Quantum Europe strategy and the EU Declaration on Quantum Technologies

[3] Draghi report (European Commission: European Political Strategy Centre, 2025), page 5.

## 3 ALIGN THE DEVELOPMENT OF QUANTUM TECHNOLOGIES WITH MOTIVATIONS OF YOUNGER GENERATIONS

Europe should put the motivations of younger generations into its Quantum Act. Europe has a strong base of highly educated people in quantum technologies, yet some leave Europe or do not realize the same developmental impact in Europe as elsewhere. Higher private investment, while important, will not automatically create the same motivational effects as in the United States. Europe should take its own motivational environments more seriously and link support instruments more directly to the aspirations of diverse younger researchers, founders, and developers. This requires combining existing mission-oriented approaches – sustainability and the moonshots – with more deliberate efforts to articulate credible and socially meaningful visions for the development of quantum technologies in Europe. In doing so, the idealistic energy of those younger generations can be effectively activated which would increase not only the number of people who want to participate in the QT development, but also the motivation of those already participating.

- **Quantum as the cool tech**

## 4 EMPHASIZE INTERNATIONAL COOPERATION AND TRADE AS A CORE EUROPEAN STRENGTH

Europe must strengthen its own quantum capabilities, but international cooperation need to play a central role in the development as well. Since current market sizes are still small and as talents, materials and knowhow are limited, the successful development of quantum technologies cannot be achieved in an autonomous manner. Instead, Europe should view selective and well-structured cooperation as a strategic asset. If Europe wants to avoid merely selling quantum technologies into markets shaped by others, and instead become the economic quantum powerhouse, it must build these relationships now. This means treating external partners not only as customers, but as actors whose needs and priorities should inform present cooperation formats. If successfully implemented, European actors will understand the needs of international markets better and be able to deliver products which are needed elsewhere, but also attract global talents, co-decide on global standards, and shape future markets. Such cooperation will not emerge automatically: partners need credible reasons to engage, and Europe's quantum technology capabilities can serve not only as assets to be protected, but also as instruments for collaboration.

- **Quantum as the economic prosperity tech**

## 5 | DEVELOP QUANTUM TECHNOLOGIES APPLICATIONS GLOBALLY AND INCLUSIVELY

Quantum technologies should be developed inclusively and in close exchange with global communities. Quantum technologies will require substantial numbers of highly trained developers, users and supporting professionals. In light of demographic changes in Europe, this makes immigration and international talent cooperation strategically relevant. Signals of closure would therefore be counterproductive. Europe should instead pursue a forward-looking quantum technology migration and cooperation agenda based on mobility, mutual learning and long-term partnership. Rather than relying on extraction alone, it should foster more equal forms of cooperation, especially with countries that have younger populations and growing talent bases. This could include collaborative programs and hackathons with partners in the Global South, designed together with European education providers and companies.

- **Quantum as the global cooperation tech**

## 6 | MAKE TECHNOLOGY ASSESSMENT A CONTINUOUS PART OF QUANTUM TECHNOLOGIES' DEVELOPMENT

Different forms of technology assessment need to become a continues part of the development of quantum technologies. Existing research suggests that negative or unintended consequences are not always visible at the outset, especially those arising from the convergence of quantum technologies with other emerging technologies, which are difficult to predict with using traditional foresight methods. Continuous risk assessment is therefore important, but a narrow focus on foreseeable risks is insufficient. It may miss genuinely unforeseen effects and problematic development paths shaped by flawed assumptions in research and innovation. Moreover, an only risk-based approach also misses the opportunity to shape the technology development from a visionary societal standpoint. The EU Quantum Act should therefore support broader and more continuous forms of assessment, including hermeneutic and real-time approaches. These approaches can map societal implications alongside technical progress and provide abilities to steer its development in a socially robust way. Technology assessment should therefore become part of strategic steering, not a corrective applied too late.

- **Quantum as the socially robust tech**

## 7 CREATE SHARED SOCIOTECHNICAL REPOSITORIES

Repositories where sociotechnical knowledge and experience is stored, need to be created. The promises of quantum technologies depend not just on the development of the hardware, software, and human ware. Instead, they depend also on their successful societal embedding, which is based on public acceptance and trust in the technology, its developers, regulators, and governance structures. Therefore, use cases and regulatory schemes need to emerge, grow, and be documented, so that they can serve as best practices for further development. For supporting this, Europe should arrive at repositories of uses case, regulation and policies relevant to quantum technologies, and create structures for developing these. Here, participatory, active involvement of the creative industries, law and policy are key, so that quantum technologies can become successfully accepted, trusted, and embedded in society beyond quantum research, engineering, and industry.

- **Quantum as the shared tech**

## 8 DEVELOP A EUROPEAN NARRATIVE FOR THE FUTURE OF QUANTUM TECHNOLOGIES

Europe should develop its own story of how quantum technologies can be shaped in a European way. As technological development is not purely deterministic, but societal systems, hopes, and dreams are co-shaping their trajectories, a genuine European approach in developing quantum technologies requires its own narrative. This narrative, based on Europe's humanistic values, as well as on the historically different cultures, would become a unique selling point for quantum technologies made in Europe and therefore support a technology development which is truly sovereign. But this cannot be formed in a top-down approach. It must emerge between the different stakeholder of a Europe's heterogenous family. Its development should therefore involve the whole variety from science, industry, policy, civil society, and the arts. If Europe wants to take these geopolitical rearrangements serious, a European narrative which leverages European strengths is key to success.

- **Quantum as Europe's sovereignty tech**

# 08 | Conclusion

This position paper offers policy makers an invitation to draw on Europe's asset of ELSA in emerging technologies. It is an invitation where we connect the reality which we acknowledge with a future that we imagine.

As ELSA scholars, we want to be clear and honest: we understand that quantum technologies are maturing, becoming more powerful, and therefore strategically relevant. We acknowledge the pressure of geopolitical shifts, and the importance of a well-articulated Quantum Act. And we recognize that everything that might seem to hinder the successful development of quantum technologies in those times is viewed as risk itself. But with this paper, we offer a different perspective. We want to acknowledge the reality and, at the same time, make an offer where the long-term benefits of an ELSA-based approach can be reaped. Because we value the ELSA-principles not only to mitigate risks. We are also convinced that including our recommendations will be beneficial for the people and the quantum technologies in Europe – socially, economically, and in terms of technological sovereignty.

We therefore have shown that ELSA is not only about mitigating risks; it is simultaneously about providing a path to socially robust and desirable technology development, and reaping its benefits. To paraphrase Mario Draghi in his report on European competitiveness: *Productivity growth can and must be linked with social inclusion*.

We have therefore delivered eight aspirational recommendations as an ELSA-offer. It is an offer, where Europe's values are used as a basis for the successful development of quantum technologies. We truly believe that if European values are incorporated in the European Quantum Act, and if its policy targets become moonshots; if the motivations of the younger generations become central; if international collaboration is taken serious and inclusive; if technology assessment is embedded as a continuous process; if sociotechnical repositories are created; and if overall a narrative for a European approach to quantum development is shaped – if the Quantum Act thereby truly focuses on Europe's strengths – then quantum technologies can be successfully and sovereignly developed in Europe.

## Acknowledgements

The ELSA policy recommendations presented in this paper were formulated solely by the authors AS, PV, and ZCS and therefore do not necessarily represent the views of the wider ELSA for quantum technologies community. Nevertheless, they draw substantially on research conducted within, and discussions held across, this community.

We are particularly grateful to the participants of the Responsible Quantum Technologies Workshop in April 2026 (ResQT 2026), where the main arguments of this paper were discussed. Their comments, and constructive suggestions provided valuable input for the further development of the paper. Additionally, we would like to acknowledge the broader community of researchers working at the intersections between ethical, legal, and social aspects (ELSA), responsible research and innovation (RRI), technology assessment, and adjacent disciplines on the one hand, and quantum technologies on the other. Their research has been instrumental in shaping this work.